\documentclass[aps,prl,preprintnumbers,showpacs,superscriptaddress]{revtex4}

\begin{document}

\title{Interplay Between Structural Randomness, Composite Disorder, and Electrical Response:
Resonances and Transient Delays in Complex Impedance Networks}


\author{R. Huang}
\affiliation{Department of Physics, Applied
Physics, and Astronomy, Rensselaer Polytechnic Institute, 110
8$^{th}$ Street, Troy, NY 12180--3590, USA}

\author{G. Korniss}
\email{korniss@rpi.edu}
\affiliation{Department of Physics, Applied
Physics, and Astronomy, Rensselaer Polytechnic Institute, 110
8$^{th}$ Street, Troy, NY 12180--3590, USA}

\author{S.K. Nayak}
\affiliation{Department of Physics, Applied
Physics, and Astronomy, Rensselaer Polytechnic Institute, 110
8$^{th}$ Street, Troy, NY 12180--3590, USA}

\begin{abstract}
We study the interplay between structural and conductivity
(composite) disorder and the collective electrical response in
random networks models. Translating the problem of time-dependent electrical response
(resonance and transient relaxation) in binary random composite networks to the
framework of generalized eigenvalues, we study and analyze the
scaling behavior of the density of resonances in these structures.
We found that by controlling the density of shortcuts (topological randomness) and/or
the composite ratio of the binary links (conductivity disorder), one
can effectively shape resonance landscapes, or suppress long
transient delays in the corresponding random impedance networks.
\end{abstract}


\date{\today}
\maketitle

\noindent

Resistor networks have been widely studied since the 70's as models
for conductivity problems and classical transport in disordered
media \cite{Kirkpatrick73}.
With the recent surge of research on complex networks
\cite{BarabREV,MendesREV},
resistor-networks and related flow models have been employed to study and explore
community structures in social networks \cite{NEWMAN04}
and to construct recommendation models for community networks \cite{Zhang_PRL2007}.
Also, resistor networks, as abstract models for network flows with a
fundamental conservation law \cite{Wu_JPA2004}, were utilized to
study transport in scale-free (SF) \cite{Andrade_PRL2005,Lopez2005,GK_PRE2007},
in a class of self-similar \cite{Makse_PNAS2007}, in small-world (SW) networks
\cite{Korniss_PLA}, and in tree structures and hierarchial lattices \cite{Antal_PRE2006}.

Complex impedance networks have been investigated to study
electrical and optical properties of two-dimensional thin films
\cite{Straley1979,Zhang1995}, and dielectric resonances of
two-dimensional regular lattice structures, lattice animals, and
other fractal clusters
\cite{Clerc1990,Clerc1996,Jonckheere1998,Wu_complex}.
In this Letter, we investigate electrical response
(resonances and delays) when {\em both} the structure and the
composition of the local conductances can be random, and we focus on the
interplay between structural and composite disorder, and response.
Random structures, in particular, random nanowire networks, can play a key role in the design and
fabrication of future electronics devices, such as
transistors or interconnects \cite{Teuscher_IEEE2006,Teuscher_Chaos2007,Cao_Nature2008,Gruner2006}.
Assessing performance and reliability of these systems requires to
understand the time-dependent intrinsic electrical response
(resonances and transient delays) of these devices which have to
switch electric currents on and off, and driven by high clock
speeds.
For example, in a random nanowire network made of
single-wall carbon nanotubes, the individual wires can be either
conductors or semiconductors (based on their individual
chiralities), resulting in links with (binary) composite disorder;
their natural composition comes with a dominance of the
semiconducting tubes \cite{Gruner2006}.
Likewise, inherent delays in electrical signal propagation can have crucial effects on processes in neuronal networks.
The compartmental-model representation of passive dendritic trees is an
$R_{1}C$-$R_{2}$ network (each compartment consists of an $R_1$
membrane leakage resistor in parallel with a capacitance $C$, and
compartments are connected with an $R_2$ junctional resistor)
\cite{Koch}. The framework employed here can be employed
to study the effects of local defects (damaged or destroyed links)
on global signal delays, ultimately governed by the structure and
link disorder in the network.

Here, we focus on the interplay between topological
randomness, conductivity disorder, and system response.
While the resonance and relaxation properties are well understood in
low-dimensional structures with conductivity (bond) disorder
\cite{Jonckheere1998}, and recently on the complete graph
\cite{Fyodorov_JPA1999,Fyodorov_2001}, to our knowledge, a similar
investigation on complex random network structures with link
conductivity disorder have not been initiated or explored.
We employ the framework applicable to binary link disorder
\cite{Straley1979,Clerc1990,Jonckheere1998,Fyodorov_JPA1999,Fyodorov_2001}.
The powerful feature of the framework is that it can be employed to
study the singularities of the electrical response associated with
any kind of binary link disorder on any graph (random {\it
L-C}, {\it RL-C}, {\it R-C}, or more complicated composite circuits,
involving two, but individually arbitrarily complicated building
blocks).

The equations governing current flows in any network can be written as \cite{Wu_JPA2004,Korniss_PLA,Wu_complex}
$\textstyle{\sum_{j}}\sigma_{ij}(V_i -V_j) = I(\delta_{is}-\delta_{it})$,
where $\sigma_{ij}$ are now the possibly complex link conductances (or admittances).
Nodes $s$  and $t$ are the nodes where a current $I$ enters and leaves the network,
respectively. The above equation can be rewritten as $\textstyle{\sum_{j}}L_{ij} V_j = I(\delta_{is}-\delta_{it})$,
where $L_{ij}=\delta_{ij}\sum_{l\neq i}\sigma_{il}-\sigma_{ij}$ is the Laplacian of the underlying graph with complex
couplings, also referred to as the admittance matrix in the present context.

For example, in {\it L-C} composite networks (or in {\it RL-C} composite
networks with weak dissipation), in order to find resonance
frequencies one can identify the non-trivial singularities, the ``zeros" of the admittance
matrix (corresponding to the ``poles" of the complex impedance matrix),
i.e., requiring that zero input current gives rise to finite
potential differences in the network \cite{Jonckheere1998,Wu_complex}.
One can also show that the zeros
of the conductivity matrix are directly related to the transient
relaxation times in {\it R-C} composite networks,
characterizing how fast the system responds to step-like on/off signals.
\cite{Clerc1990,Jonckheere1998}. For binary composite networks, the conductance
disorder of existing links in the structure is characterized by a
single parameter (composite ratio) $q$, such that $\sigma_{ij}=\sigma_{1}$ with
probability $q$, and $\sigma_{ij}=\sigma_{2}$ with probability $(1-q)$
(and obviously, $\sigma_{ij}=0$ if nodes $i$ and $j$ are not
connected).
For example, for an {\it L-C} composite network, $\sigma_{1}=i\omega C$, $\sigma_{2}=(i\omega L)^{-1}$,
while for an {\it R-C} composite network, $\sigma_{1}=i\omega C$, $\sigma_{2}=R^{-1}$.
Hence, for resonance condition in {\it L-C}
networks (and for relaxation times for {\it R-C} networks), one searches
for the nontrivial solutions of
$\textstyle{\sum_{j}}L_{ij}(\omega)V_j = 0$, or
${\mathbf  L}(\omega){\mathbf  V} = {\bf 0}$
in a more compact notation. Then for any fixed graph and any realization of the binary link
disorder, one can rewrite the above expression
for the resonance condition (or to extract transient relaxation times) \cite{Fyodorov_JPA1999,Fyodorov_2001},
\begin{equation}
({\bf H}-\lambda {\bf \Gamma}){\bf V} = {\bf 0} \;,
\;\;\;\;\;\;\;\;\;\;
\lambda=\frac{\sigma_{1}+\sigma_{2}}{\sigma_{1}-\sigma_{2}}\;.
\label{generalized_eigenvalue}
\end{equation}
Here, $H_{ij}=\delta_{ij}\sum_{l\neq i}h_{il}-h_{ij}$, where
$h_{ij}$$=$ $-1$, $+1$, $0$ if $\sigma_{ij}$$=$ $\sigma_{1}$, $\sigma_{2}$, $0$, respectively.
Similarly, $\Gamma_{ij}=\delta_{ij}\sum_{l\neq i}h_{il}^2-h_{ij}^2$,
is just the (topological) network Laplacian of the underlying graph
($h_{ij}^2=1,0$ is obviously the adjacency matrix of the network).
The expression for $\lambda$ in Eq.~(\ref{generalized_eigenvalue})
establishes the connection between the generalized eigenvalues and
the resonance frequencies $\omega_j$ of {\it LC}, or the transient relaxation times $\tau_j$ of
{\it RC} composite networks \cite{Clerc1990,Jonckheere1998},
\begin{equation}
\omega_j=\frac{1}{\sqrt{LC}} \sqrt{\frac{1+\lambda_j}{1-\lambda_j}}\;,
\;\;\;\;\;\;\;\;
\tau_j = RC \frac{1-\lambda_j}{1+\lambda_j}\;.
\label{omega_tau}
\end{equation}
Hence the above generalized eigenvalue problem
Eq.~(\ref{generalized_eigenvalue}), where ${\bf \Gamma}$ is real
symmetric and nonnegative and ${\bf H}$ is real symmetric, provides
a framework to identify the resonance frequencies (density of
resonances in the large $N$ limit) or relaxation times in the
respective binary composite networks.
It is also clear from the above framework that the resonance (and
relaxation) spectrum (except from pathological cases) is {\em
independent} of the choice of nodes where the current enters and
leaves the system, thus, they represent {\em intrinsic
characteristics} of the network
\cite{Clerc1990,Jonckheere1998,Wu_complex}.

In what follows, for brevity, we use the ``resonance" terminology in
composite networks, and will also refer to $\lambda$
as ``frequency". The eigenvalues of the above system
always fall in the $[-1,+1]$ interval,
$\lambda_{1}\leq\lambda_{2}\leq\ldots\leq\lambda{_N}$, with true
resonances corresponding to $-1<\lambda_{j}<1$.
We focused on two important observables, the density of resonances $\rho(\lambda)$
and number of resonances {\em per node} $\overline{\rho}$
\cite{Clerc1990,Jonckheere1998,Fyodorov_JPA1999,Fyodorov_2001}
\begin{equation}
\rho(\lambda)=\frac{1}{N}\sum_{\alpha=1}^{n_R}\delta(\lambda-\lambda_{\alpha})\;,
\;\;\;\;\;\;\;\;
\overline{\rho}=\int\rho(\lambda)d\lambda = \frac{n_R}{N}\;,
\label{dens_res}
\end{equation}
where $n_R$ is the total number of true resonances (not associated
with $\lambda_{j}=\pm1$).
In this work, we determined the spectrum of the generalized eigenvalue
problem Eq.~(\ref{generalized_eigenvalue}) numerically, and constructed the above
observables by averaging over 10,000 realizations (1,000 realizations for the largest system size)
of both structural and composite disorder.
\begin{figure}[t]
\vspace*{-0.50truecm}
\centering
\includegraphics{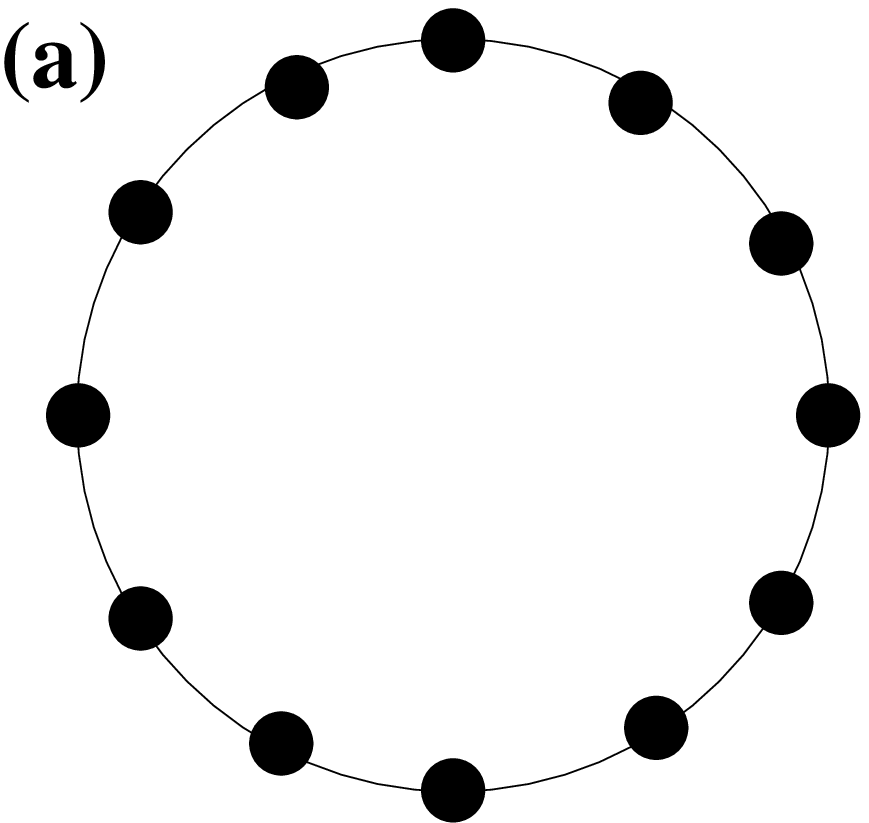}
\includegraphics{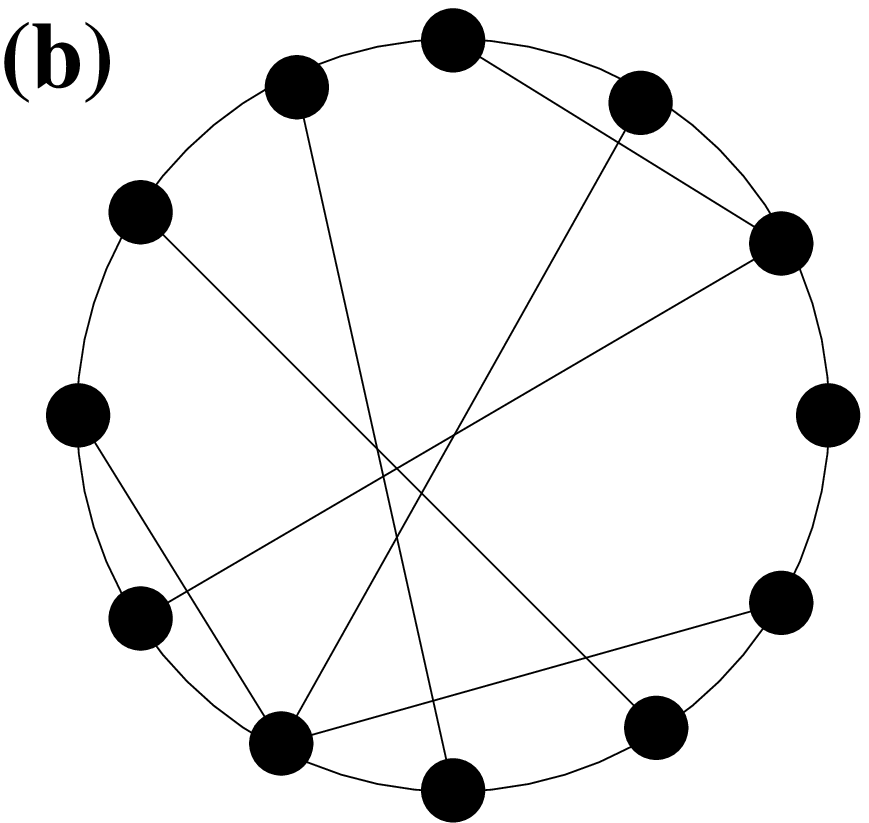}
\includegraphics{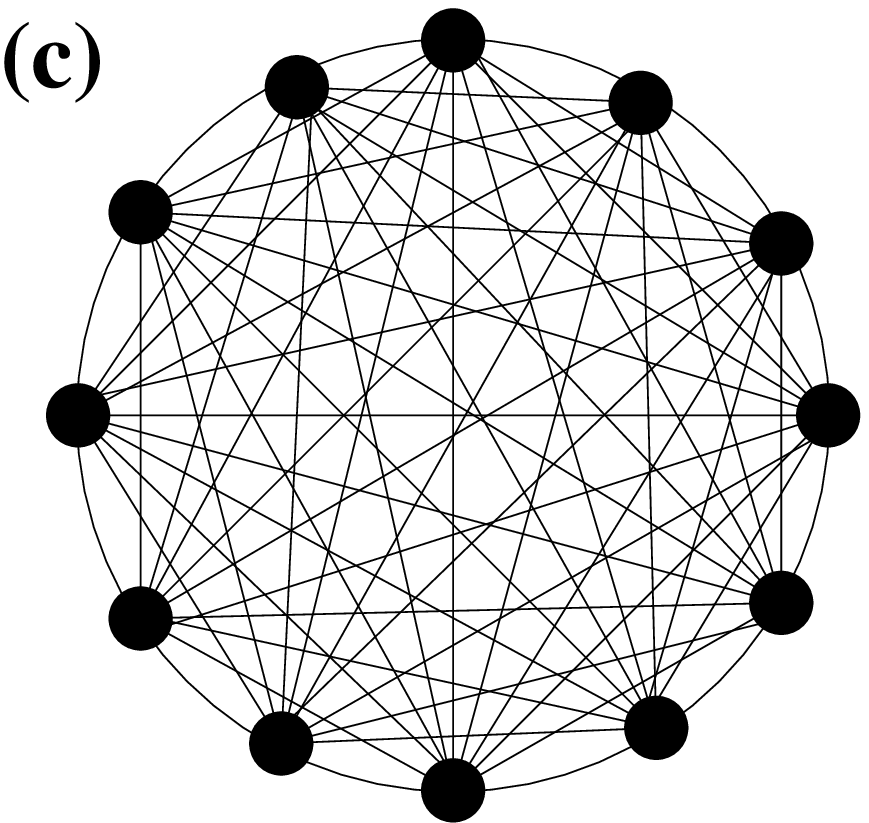}
\vspace*{3.50truecm}
\caption{
Network structures:
(a) one-dimensional ring;
(b) small-world network with random shortcuts added between nodes on a ring (1d SW) \cite{Newman_Watts};
(c) complete graph (each node is connected to all others).
In binary composite networks, each existing link has a complex conductance $\sigma_1$ or $\sigma_2$
with probability $q$ and $1$$-$$q$, respectively.}
\label{network_config}
\vspace*{-0.10truecm}
\end{figure}

Before studying random structures with binary composite disorder,
we recall two known extreme cases: the one-dimensional ring
[Fig.~\ref{network_config}(a)] and the complete graph
[Fig.~\ref{network_config}(c)], both with the same composite (binary link) disorder.
For a one-dimensional ring, there is a single resonance frequency
\cite{Wu_complex} (which can also be obtained via elementary
considerations). More specifically, $n_R=1$ and the frequency is distributed binomially
about $\langle\lambda\rangle=2q-1$.
Thus, in the large-$N$ limit, the density of resonances Eq.~(\ref{dens_res}) approaches a Gaussian
distribution with the above mean and vanishing width, i.e., a delta
function. For example, for $q$$=$$1/2$,
$\rho(\lambda)\simeq\frac{1}{N}\delta(\lambda)$ and $\overline{\rho}=\frac{1}{N}$.

In the other extreme case where all nodes are connected to all others
(i.e., the complete graph), using a path-integral approach \cite{Fyodorov_JPA1999,Fyodorov_2001,Staring}
in the large-$N$ limit, Fyodorov obtained (without loss of generality, for $q$$=$$1/2$) that
$\rho(\lambda) \stackrel{N\to\infty}{\longrightarrow}\delta(\lambda)$ and
$\overline{\rho} \stackrel{N\to\infty}{\longrightarrow} 1$
i.e., the total number of resonances approaches the number of nodes, but
they are all narrowly centered about the same frequency (becoming
fully degenerate as $N$$\to$$\infty$).
\begin{figure}[t]
\vspace*{2.0truecm}
\centering
\includegraphics{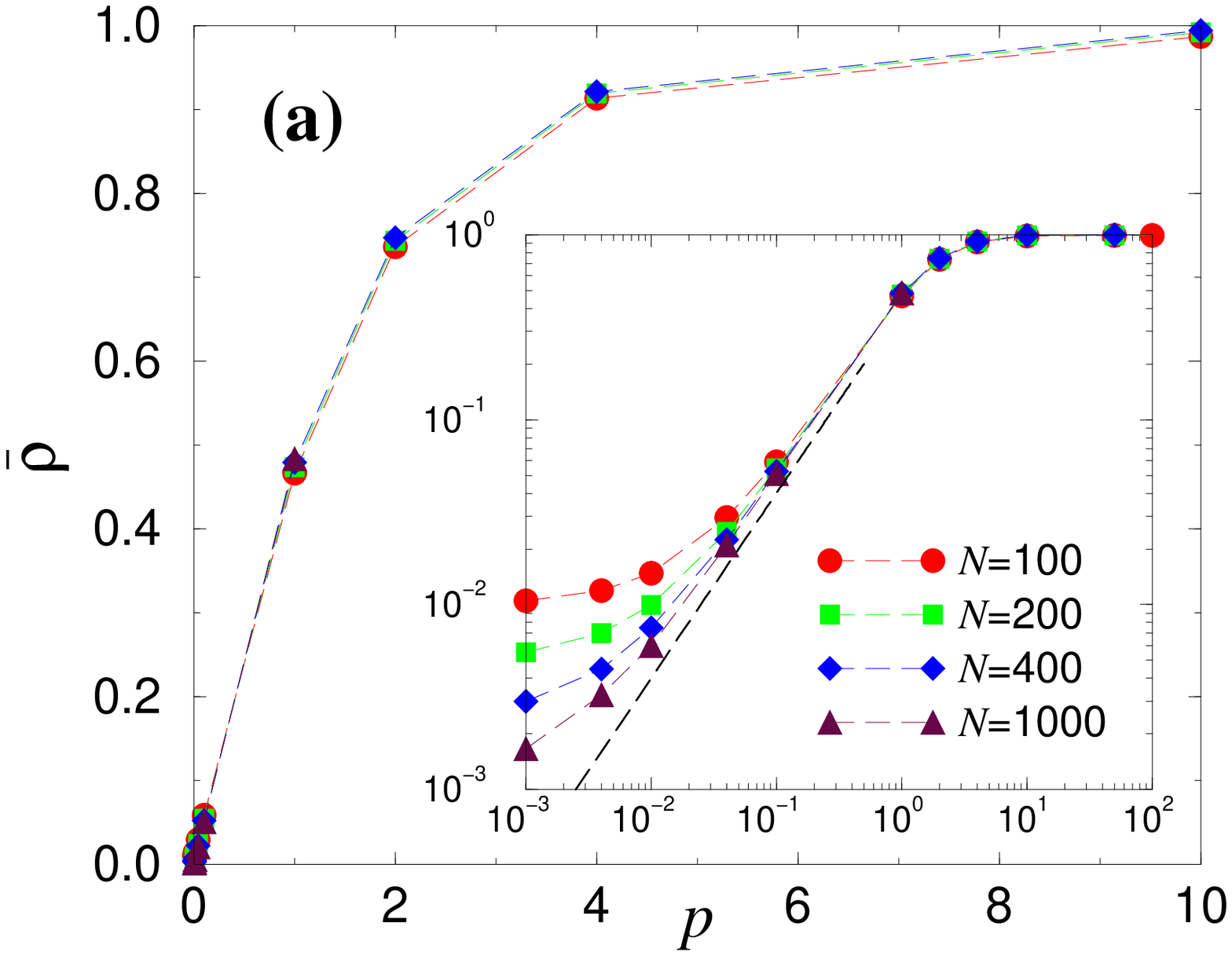}
\includegraphics{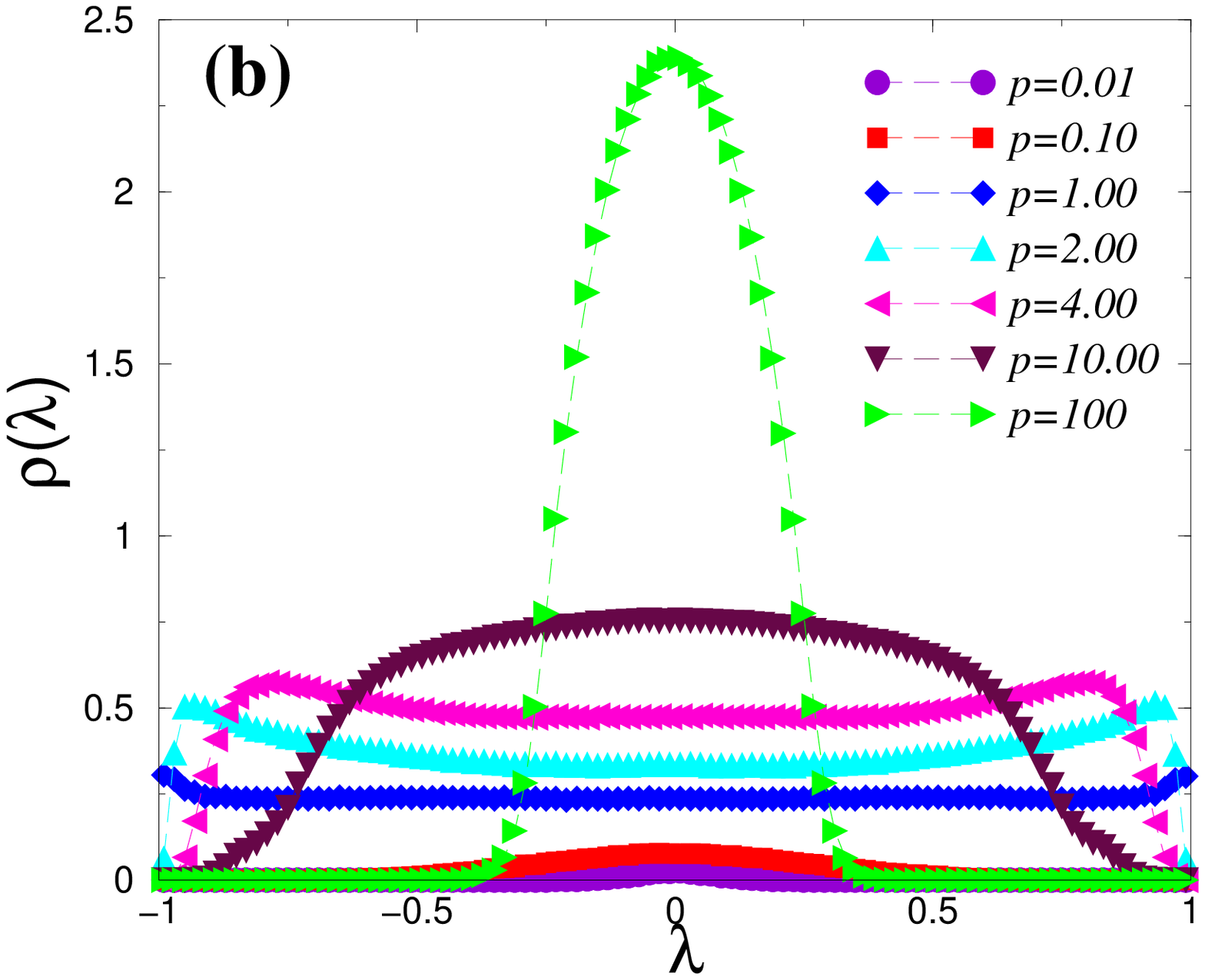}
\includegraphics{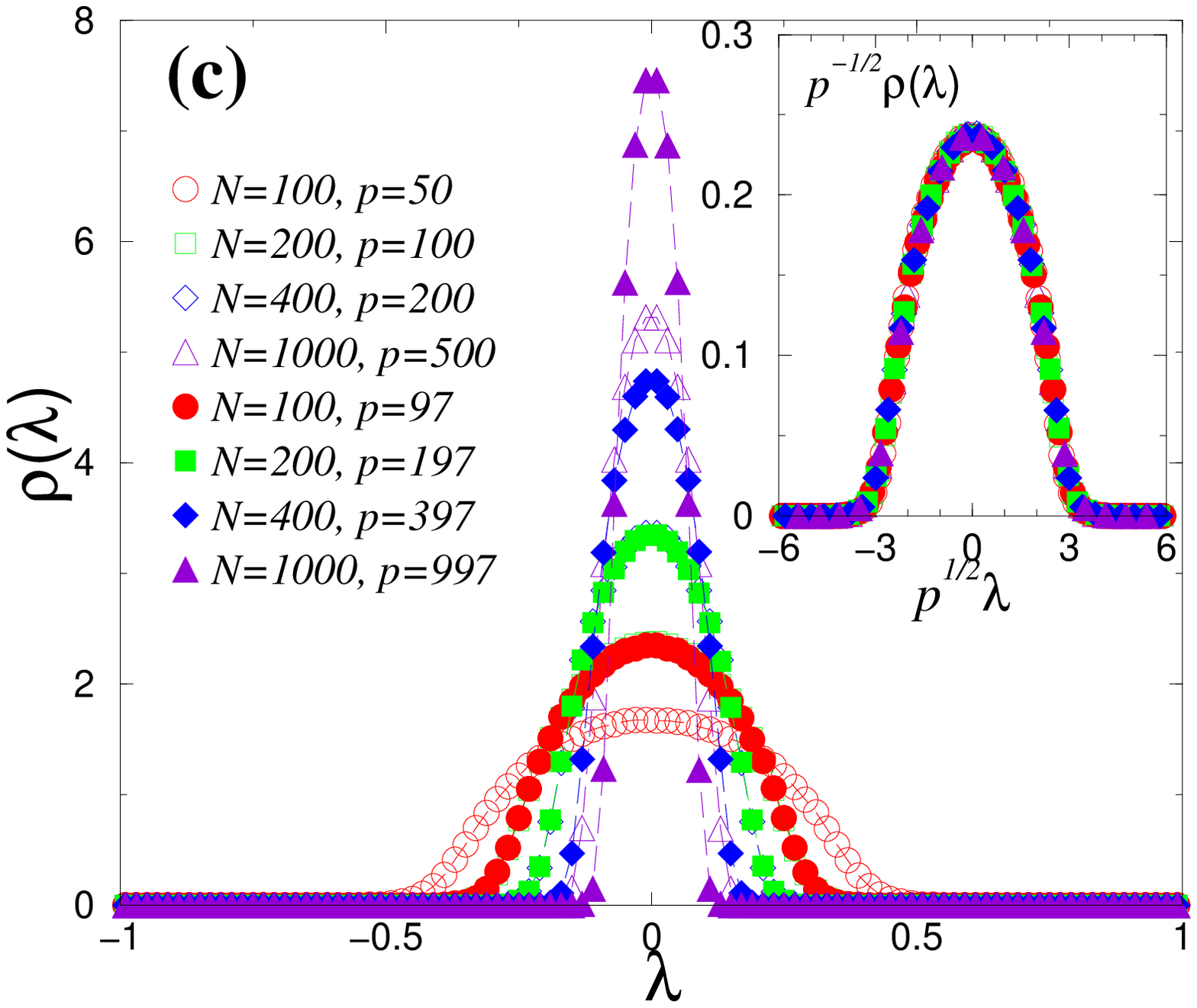}
\vspace*{2.80truecm}
\caption{ (a) Average number of resonances per
node vs density of shortcuts $p$ in a 1d SW network with composite
ratio $q$$=$$1/2$. The inset shows the same data on log-log scales,
with the straight dashed line indicating the asymptotic large-$N$
behavior, $\overline{\rho}\sim p$.
(b) Density of resonances for fixed number of nodes $N$$=$$1000$ with $q$$=$$1/2$,
and for different values of $p$.
(c) Density of resonances in the large-$p$
regime for $q$$=$$1/2$, for various system sizes. The inset shows
the scaled plot of the same data, $\rho(\lambda)/p^{1/2}$ vs
$p^{1/2}\lambda$. These and the following plots all show ensemble
averages over 10,000 network- and composite-disorder realizations
(1,000 for the largest system size).}
\label{SW_resonances_1}
\vspace*{-0.10truecm}
\end{figure}

\begin{figure}[t]
\vspace*{2.00truecm}
\centering
\includegraphics{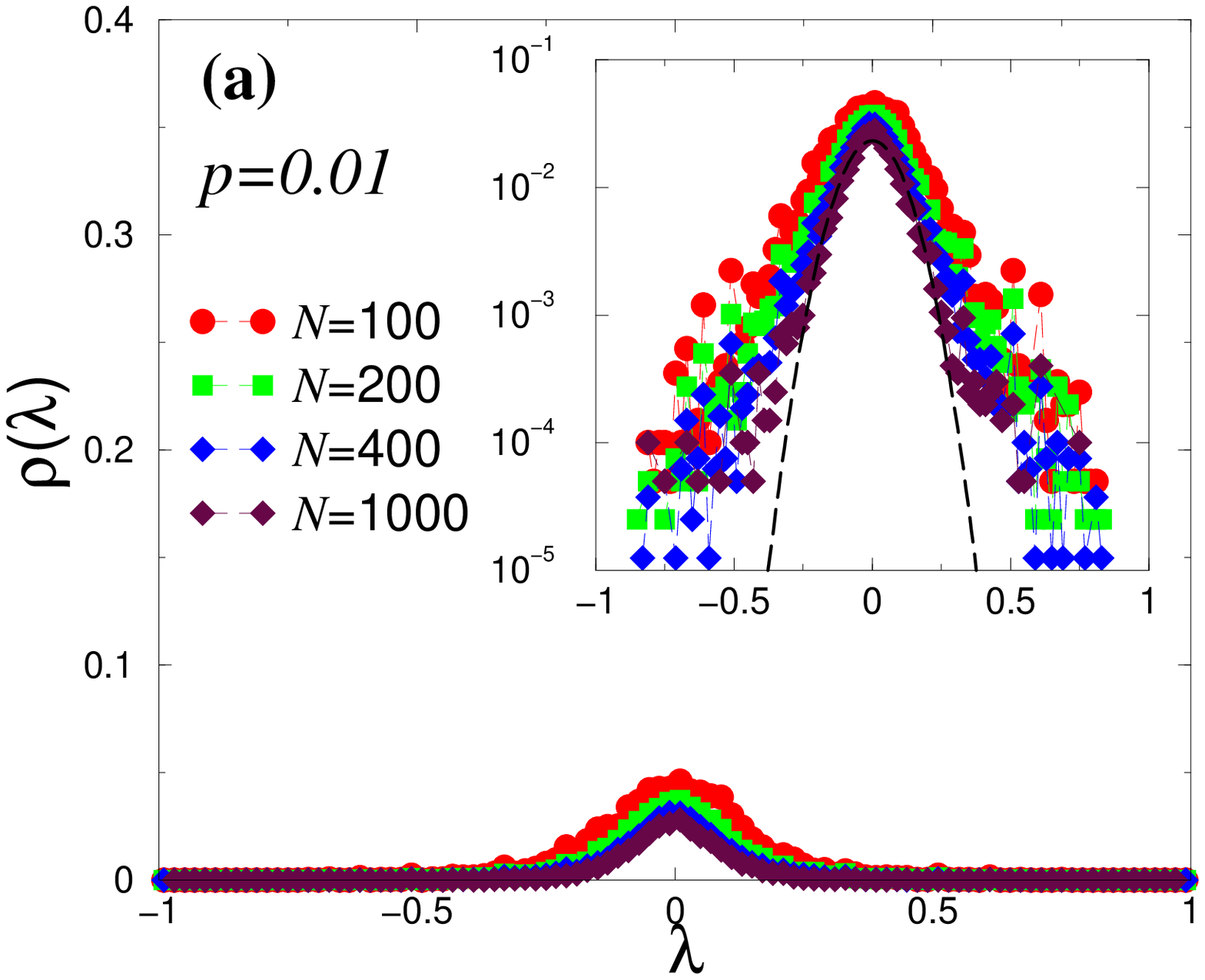}
\includegraphics{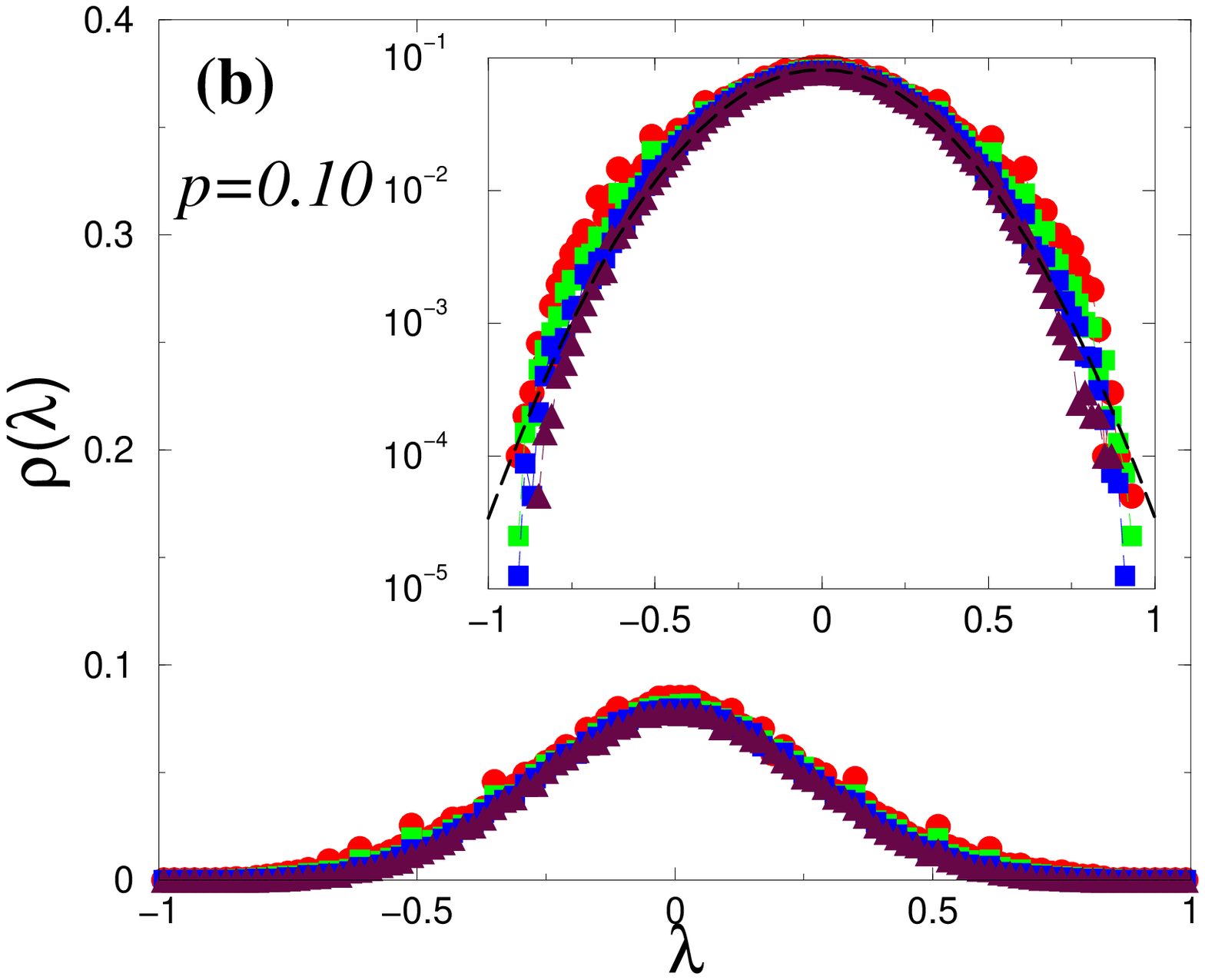}
\includegraphics{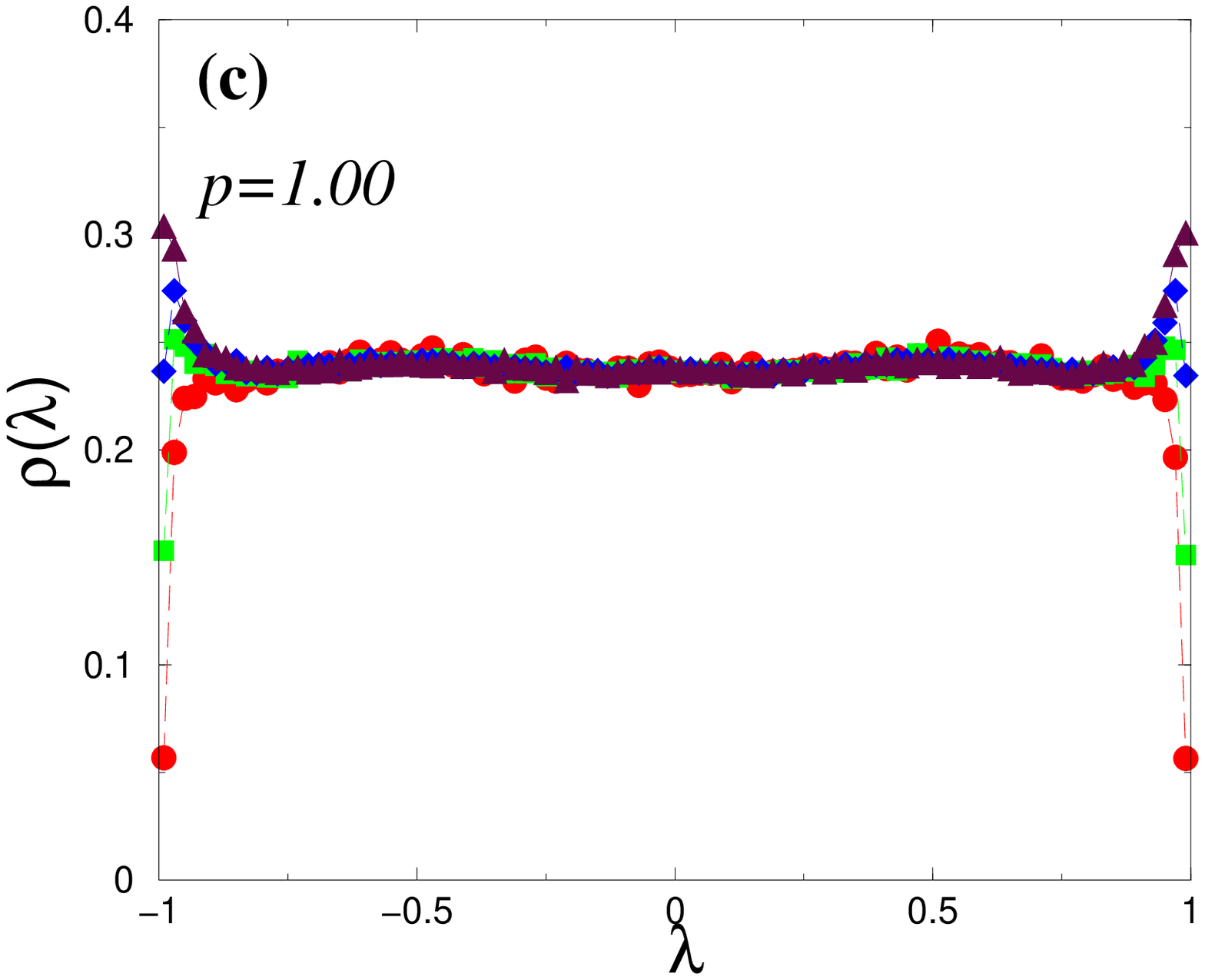}
\vspace*{2.80truecm}
\caption{
Density of resonances in 1d SW
networks for different system sizes, for composite ratio
$q$$=$$1/2$, and for (a) $p$$=$$0.01$, (b) $p$$=$$0.10$, and (c)
$p$$=$$1.00$. System sizes and the corresponding symbols in all
three panels are the same as in (a). The insets in (a) and (b) show
the same data on lin-log scales and a Gaussian fit to the largest
system size around the center (dashed curve).}
\label{SW_resonances_2}
\end{figure}

Now, we consider small-world (SW) networks \cite{WATTS98} as random structures,
where random shortcuts were {\em added} to a one-dimensional ring
(1d SW) \cite{Korniss_PLA,Newman_Watts}
[Fig.~\ref{network_config}(b)], resulting in an average number of
random shortcuts per node $p$. For comparison with the previous two
extreme cases, we show results for the same composite ratio
$q$$=$$1/2$. The results show that for any nonzero value of $p$, the
number of resonances per node will approach a nonzero
$\overline{\rho}>0$ value in the large-$N$ limit, as opposed to the
pure 1d ring where it vanishes as $1/N$
[Fig.~\ref{SW_resonances_1}(a)]. Further, as the number of random
links per node $p$ increases, $\overline{\rho}$ increases
monotonically, and the density of resonances initially [$0<p\le
{\cal O}(1)$] widens; at around $p\sim{\cal O}(1)$, the spectrum
becomes {\em extended} [Fig.~\ref{SW_resonances_1}(b)].
As we further increase $p$, the number of resonances per node
continues to increase monotonically as a function $p$, quickly
``saturating" to its maximum value $\overline{\rho}=1$
[Fig.~\ref{SW_resonances_1}(a)], while the density of resonances
becomes progressively centered about $\lambda$$=$$0$
[Fig.~\ref{SW_resonances_1}(b,c)], eventually converging to a
delta-function (if both $p$$\to$$\infty$, $N$$\to$$\infty$). Indeed,
one can recall for the complete graph, that the average number of
resonance per node approaches $\overline{\rho}=1$, but all
frequencies are centered about the same value
\cite{Fyodorov_JPA1999,Fyodorov_2001}. Note that in both the
low shortcut density [$0<p\ll {\cal O}(1)$,
Fig.~\ref{SW_resonances_2}(a,b)] and the high shortcut density [$p\gg
{\cal O}(1)$ (not shown)] regimes, for fixed $p$, the density of
resonances becomes independent of the size of the network for large
$N$. Finite-size effects are very strong, however, for $p\sim{\cal
O}(1)$, in particular in the low- and high-frequency regime
Fig.~\ref{SW_resonances_2}(c)].
Further analysis in the high-connectivity [$p\gg {\cal O}(1)$]
regime also reveals that the limit density of resonances has the
scaling form $\rho(\lambda)=p^{1/2}\phi(p^{1/2}\lambda)$
[Fig.~\ref{SW_resonances_1}(c), and inset]. This scaling form,
valid for all $p\gg {\cal O}(1)$, is identical to the one found for
regular long-range connectivity graphs \cite{Staring}, including the
limit of complete graph ($p$$\to$$N$) \cite{Fyodorov_note}.

Next, we provide more details for the low shortcut-density regime (also referred to as the SW regime), $0<p\ll{\cal
O}(1)$. The scaling of the number of resonances per node
$\overline{\rho}$ in this regime [Fig.~\ref{SW_resonances_1}(a)],
can be extracted from the finite-size behavior, typical in 1d SW
networks \cite{Korniss_PLA,Newman_Watts}.
Since the number of random links per node (density of shortcuts) is
$p$, the typical (Euclidean) distance between nodes with shortcuts
emanating from them scales as $\xi\sim p^{-1}$. Thus, for $N\ll\xi$
($Np\ll 1$), there are no random links in almost any realization of
the network, and the resonance structure will essentially be
identical to that of the pure ring. A crossover, governed by the
emerging SW structure, can be expected when $N\gg\xi$ ($Np\gg 1$).
Thus, in the SW regime, $0<p\ll{\cal O}(1)$, for arbitrary $N$, the
above crossover behavior of $\overline{\rho}(N,p)$ can be expressed
in terms of $N$ and the scaled variable $x=Np$ with the help of a
scaling function $f(x)$, such that
\begin{equation}
\overline{\rho}(N,p) =\frac{1}{N}f(Np)\;,
\label{SW_scaling}
\end{equation}
where $f(x)$$\sim$${\rm const.}$ for $x$$\ll$$1$, while
$f(x)$$\sim$$x$ for $x$$\gg$$1$. Thus, in the large network-size
limit ($N$$\to$$\infty$), for {\em small} values of $p$,
$\overline{\rho}\sim p$,
i.e., the number of resonance per node increases linearly with the
average number of random links per node, as can be seen in
Fig.~\ref{SW_resonances_1}(a) (inset). Our numerical results also
suggest (although with considerable finite-size effects for finite
networks) that for any {\em fixed} $0<p\ll 1$ value, $\rho(\lambda)$
approaches a system-size-independent limit density and obeys the
scaling form
$\rho(\lambda)= \overline{\rho}(p) p^{-1/2}\Psi(\lambda/p^{1/2}) \sim p^{1/2}\Psi(\lambda/p^{1/2})$.
Further, the scaling function $\Psi(s)$ is reasonably well approximated by
$\sim e^{-c s^2}$ in the vicinity of the center as $N$$\to$$\infty$, i.e., the density of resonances approaches a Gaussian shape
[Fig.~\ref{SW_resonances_2}(a,b)].

\begin{figure}[t]
\vspace*{2.00truecm}
\centering
\includegraphics{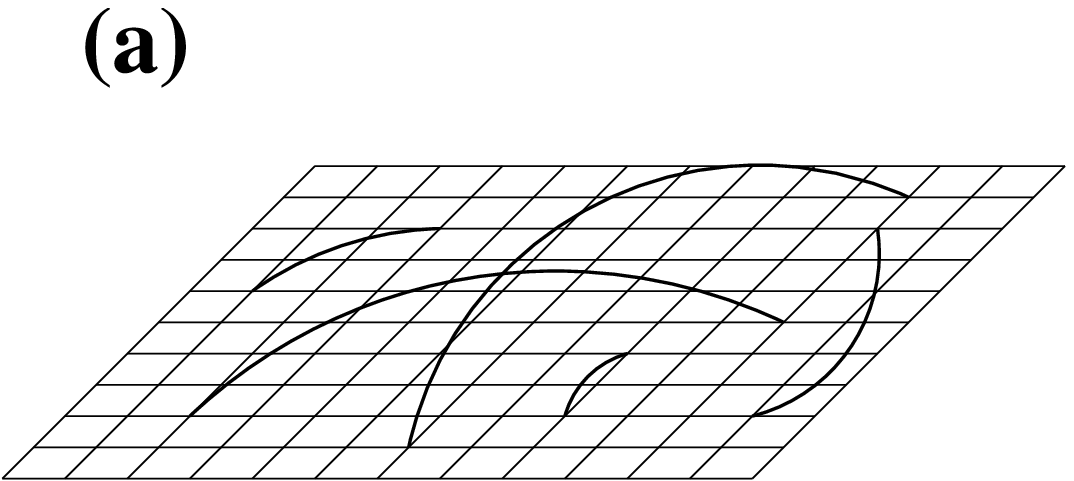}
\includegraphics{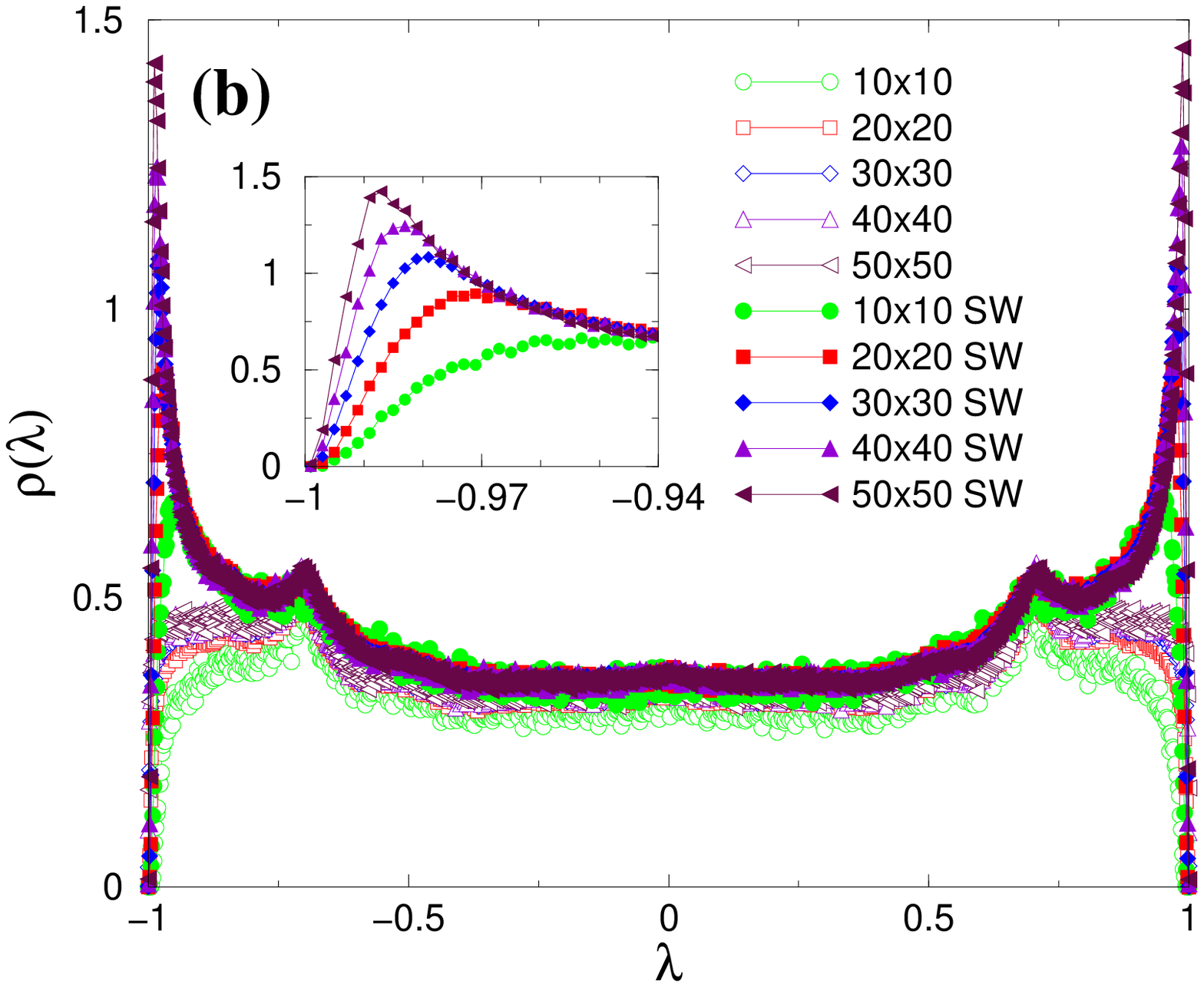}
\includegraphics{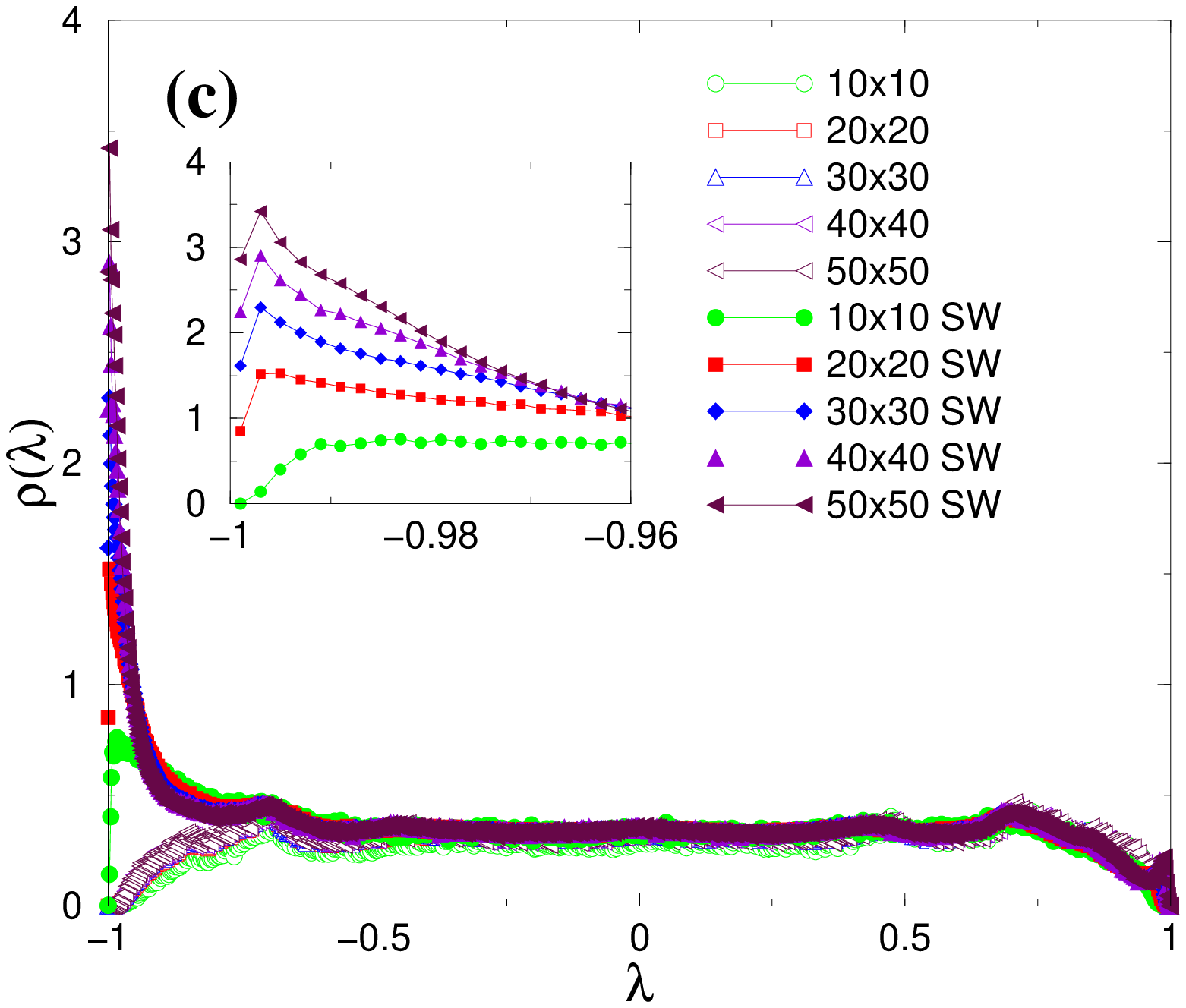}
\vspace*{2.70truecm}
\caption{
(a) Schematic plot of a 2d SW graph.
(b) Density of resonances in regular 2d ($p$$=$$0$, open symbols) and power-law-suppressed 2d SW networks
($p$$=$$1.00$, $\alpha$$=$$1.00$, solid symbols) for different system sizes ($N$$=$$L$$\times$$L$),
and for composite ratio $q$$=$$1/2$. The insets show the same data enlarged
for the 2d SW network in the vicinity of one of the peaks.
(c) The same plot as (b) for composite ratio $q$$=$$2/3$.}
\label{2dSW_resonances}
\vspace*{-0.1truecm}
\end{figure}

To model more complicated spatially-embedded random structures [Fig.~\ref{2dSW_resonances}(a)], we
considered when {\em both} the value of the complex link
conductivity and the probability to have a link between two nodes
can depend on Euclidean distance between the two nodes it connects.
From elementary length scale considerations, for the
distance-dependent link conductivity, one has $\sigma_{ij}\sim
1/d_{ij}$, where $d_{ij}$ is the Euclidean distance between nodes
$i$ and $j$ (assuming uniform ``wire" cross-sections)
[Eq.~(\ref{generalized_eigenvalue}) easily generalizes to this
case.] Whereas the probability of having a link (shortcut) between
node $i$ and $j$, can also be suppressed e.g., $p_{ij}\sim
1/d_{ij}^{\alpha}$ (power-law-suppressed SW networks due to
``wiring"-cost considerations or topological constraints
\cite{wiring_cost,KHK05}).
In Fig.~\ref{2dSW_resonances}(b), we show the resonance spectrum
of a {\em two-dimensional} power-law-suppressed SW network (2d SW) with open boundaries
with $\alpha$$=$$1$ and $p$$=$$1.00$ (random shortcuts with
distance-dependent conductivities were {\em added} on top of a
two-dimensional regular ``substrate" [Fig.~\ref{2dSW_resonances}(a)]),  with composite  ratio
$q$$=$$1/2$, together with the known results
\cite{Clerc1996,Jonckheere1998} of the regular two-dimensional
topological structures with the same composite disorder. For regular
two-dimensional structures, in the large-system size limit,
$\overline{\rho}\sim{\cal O}(1)$ and the spectrum is known to be
{\em extended} \cite{Clerc1996,Jonckheere1998}. The addition of
distance-dependent shortcuts, however, strongly modifies the density
of resonances in the vicinity of $\lambda$$=$$\pm1$ (strong peaks
for low and high frequencies). Further, the structure of the peaks
do not approach a limit density in that region, but diverge with
system size (with $p$ fixed). An analogous plot for an asymmetric
link disorder with $q$$=$$2/3$ [Fig.~\ref{2dSW_resonances}(c)]
shows strong (diverging) peaks only in the
small-frequency regime [also translating to large transient
relaxation times or delays in {\it RC} networks
Eq.~(\ref{omega_tau})]. Our analyses also indicate that the main
qualitative features (articulated peaks for low and/or high
frequencies) of structures with distance-dependent shortcuts prevail
for a range of $\alpha$, $0$$\leq$$\alpha$$\leq$$\alpha_c$$\approx$$2\pm0.5$.


In summary, we have shown that in random composite networks, by
controlling the density of shortcuts $p$ (topological randomness)
and/or the composite ratio $q$ of the binary links (conductivity
disorder), one can effectively shape the resonance landscape, or suppress long
transient delays in electrical signal propagation.
Here, we have highlighted the interplay between structural and
composite (conductivity) disorder and the collective electrical
response in spatially-embedded random networks models. The
electrical response of more realistic off-lattice random
structures, embedded in two- and three-dimensions, reflecting
relevant wiring cost and topological constraints
\cite{wiring_cost,Teuscher_IEEE2006,Teuscher_Chaos2007} will be
considered in future works. A detailed analyses on such structures
will help one understand electrical response (resonances and signal
delays) in complex materials and biological networks.

This work was supported in part by NSF DMR-0426488. R.H. and S.K.N
were also supported in part by the Focus Center, NY at RPI.


\end{document}